\documentclass[journal]{vgtc}                

\let\chapter\section
\usepackage{graphicx}
\usepackage{times}
\usepackage[ruled, vlined, linesnumbered]{algorithm2e}
\usepackage[numbers]{natbib}
\usepackage{enumitem}
\usepackage{amsmath}
\usepackage{amsthm}
\usepackage{mathrsfs}
\usepackage{amssymb}
\usepackage{booktabs}
\usepackage{pifont}

\newcommand{\argmax}{\operatornamewithlimits{argmax}}

\graphicspath{{./figures/}}
\vgtccategory{Research}

\vgtcinsertpkg

\title{Interactive and Iterative Discovery of Entity Network Subgraphs}

\author{Hao Wu, Maoyuan Sun, Jilles Vreeken, Nikolaj Tatti, Chris North, Naren Ramakrishnan}

\abstract{
	Graph mining to extract interesting components has been studied in various
	guises, e.g., communities, dense subgraphs, cliques. However, most existing
	works are based on notions of frequency and connectivity and do not capture
	subjective interestingness from a user's viewpoint. Furthermore, existing
	approaches to mine graphs are not interactive and cannot incorporate user
	feedbacks in any natural manner. In this paper, we address these gaps by
	proposing a graph maximum entropy model to discover surprising connected
	subgraph patterns from entity graphs. This model is embedded in an
	interactive visualization framework to enable human-in-the-loop,
	model-guided data exploration. Using case studies on real datasets, we
	demonstrate how interactions between users and the maximum entropy model
	lead to faster and explainable conclusions.  
} 

\teaser{
 \centering
 \includegraphics[width=16cm]{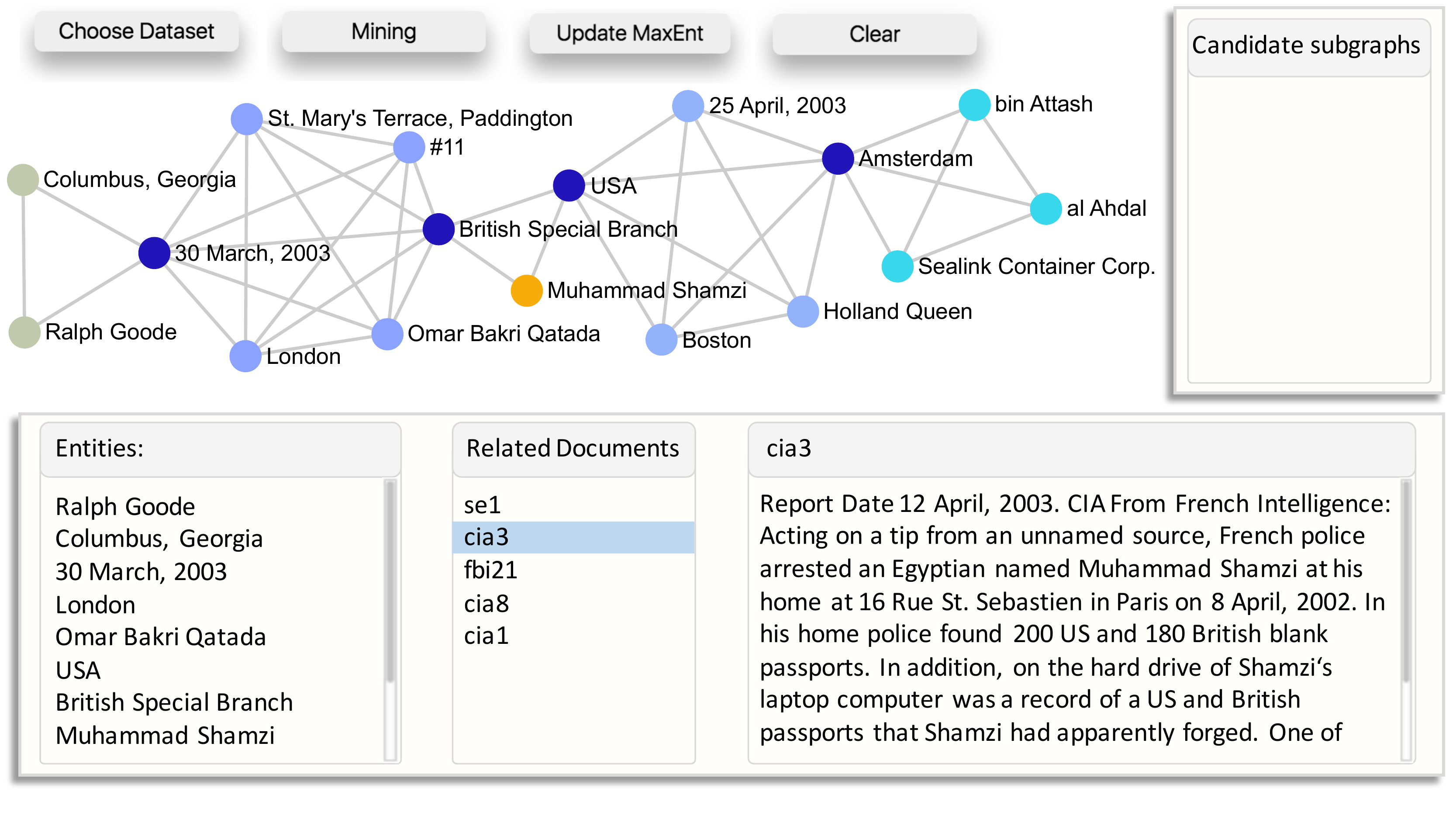}\vspace{-0.1in}
  \caption{An overview of the user interface for the proposed interactive and
  iterative visualization framework to discover entity network subgraphs. A
  connected subgraph pattern identified from an intelligence analysis dataset is
  displayed in this example. The involved entities and corresponding
  intelligence documents are listed below the displayed pattern.}
  }

\begin{document}


\firstsection{Introduction}

\maketitle

Knowledge discovery from graphs is a crutial task that arises from many
reasearch and application domains, e.g.\ social network analysis, biological
knowledge discovery, and storytelling from unstructured text documents. As the
fast development of Internet technology and social media, social networks like
Facebook and Twitter have drawn much attention in both industry and academic
community. Identifying interesting group structures, e.g.\ communities, in such
social networks can help sociologists to understand people's social behaviors in
the virtual world and compare with that of the real world. From an industrial
perspective, subgraph patterns in such social network can help companies deliver 
their advertisements more precisely and recommand their products and services to
potential clients over the Internet. In biology domain, questions like ``how do
these pathways interact and influence with each other in bological pathway
networks?'' are typical challenges faced by biologists in their research.
Similarly, in storytelling from text datasets, analysts are always interested in
interactions between various types of entities, e.g.\ persons and locations, in
entity networks, which could lead the analysts to discover the plots or even
conspiracies hidden behind phenomenons.

To address these challenges, a plenty of graph mining algorihms which extract
local patterns, such as communities~\cite{Fortunato201075}, dense
subgraphs~\cite{Lee2010} and cliques, e.g.~\cite{Eppstein2010}, from networks
have been proposed and studied in the recent decade. However, most of such
existing works focus on investigating the notions of density or connectivivty
to indentify the subgraph patterns and do not capture the subjective
interestingness from a user's perspective. Moreover, except for few works, e.g.\
Apolo~\cite{Chau:2011:AMS:1978942.1978967}, most of the existing graph mining
approaches are purely algorithmic without involving any algorithm-user
interactions, which ignores the user's important feedbacks.

To overcome such drawbacks, in this paper, we fill these gaps by proposing a
graph Maximum Entropy (MaxEnt) model together with a subjective interestingness
measure to discover interesting connected subgraph patterns from entity graphs.
By designing a greedy heuristic algorithm that works together with the proposed
graph MaxEnt model, we achieve the same goal of automatically discovering
subgraph patterns from graphs compared to the traditional graph mining
algorithms. In additional, by embedding the graph MaxEnt model into an
interactive visualization framework, we enable the iterative, human-in-the-loop,
model-guided data exploration. The key point we would like to emphasize here is
that the ultimate objective of knowledge discovery is not to extract an unique
answer from the dataset but rather to guide domain experts into deeper
consideration and understanding of key process elements.

Our contributions in this paper are:
\begin{enumerate}[itemsep=-1pt,topsep=-1pt]
	\item We derive and present the formalization of the graph MaxEnt model, and
		define a criterion that measures the interestingness of local subgraph
		patterns based on the MaxEnt model.
	\item We design a greedy heuristic algorithm to automatically discover the
		interesting connected subgraph patterns from entity graphs.
	\item By integrating the graph MaxEnt model with a visualization framework,
		we enable the model-guided, interactive and iterative data exploration
		and knowledge discovery over graphs.
	\item Using the results and a use case study on real world intelligence
		datasets, we demonstrate how our proposed graph MaxEnt model and
		visualization framework are adopted to analyze real world datasets. In
		particular, we show how our approach supports human-in-the-loop
		knowledge discovery, and leads the analysts to uncover the hidden plots
		of the datasets.
\end{enumerate}

\section{Preliminaries}
\label{sec:pre}

In this section, we will introduce some preliminary concepts that will be
useful and helpful to understand our proposed algorithm in the rest of this
paper.

\subsection{Graph Notations}
\label{sec:pre_graph}

A graph is usually defined as an ordered tuple $G = (V, E)$, where $V$
represents a set of vertices in the graph, and $E$ denotes a set of edges that
connect vertices in $V$. Usually, a tuple of two vertices that an edge connects
is used to represent this edge. A graph can be further classified as directed
graph or undirected graph depending on whether the direction of edges matters.
If we let $v \in V$ denote a single vertex in the graph, in the directed
graph, an edge $(v_i, v_j)$ from $v_i$ to $v_j$ is different from the edge
$(v_j, v_i)$ from $v_j$ to $v_i$. However, in the undirected graph, they are the
same. In this paper, we will focus on the directed graph when we describe the
proposed model since the undirected graph can be seen as a special case of
the directed graph, e.g.\ if an edge exists between vertices $v_i \in V$ and
$v_j \in V$, both edges $(v_i, v_j)$ and $(v_j, v_i)$ belong to the edge set
$E$. Thus, in the rest of the paper, all the graphs mentioned will refer to the
directed graph unless specified.

A subgraph of $G$ is a graph $G^{\prime} = (V^{\prime}, E^{\prime})$ such that
$V^{\prime} \subseteq V$ and $E^{\prime} \subseteq E$. We use $G^{\prime}
\subseteq G$ to represent the subgraph relationship. Among all the possible
subgraphs of $G$, a clique $C = (V_c, E_c)$ is a special type of subgraph such
that $\forall v_i, v_j \in V_c$ and $v_i \neq v_j$, the edge $(v_i, v_j) \in
E_c$. In other words, a clique is a subgraph that is complete. A maximal clique
is a clique that cannot be extended by adding more vertices into it.

In terms of the connected subgraph, we refer to a tuple of ordered subgraphs
$\mathit{CS} = (G_1, G_2, \ldots, G_n)$ such that the adjacent subgraphs
$G_i$ and $G_{i + 1}$ share at least one common vertex $v$. In this case, we say
the common vertex $v$ connect the subgraph $G_i$ and $G_{i + 1}$.

\subsection{MaxEnt Models}
\label{sec:pre_maxent}

The Maximum Entropy (MaxEnt) principle~\cite{jaynes:59:maxent} has drawn much
attention in the pattern mining community recently, especially in the area of
discovering subjectively interesting patterns. The concept of entropy is
originated from the information theory~\cite{Shannon:2001:MTC:584091.584093}. In
the context of data mining, entropy is adopted to measure how certain a model is
about the data. Lower entropy indicates the model is quite certain about the
data it models. It would be perfect if we find a low entropy model where the
model summarizes the majority information conveyed by the data we are modeling.
However, the given prior information about the data is usually quite limited in
practice. Inferring a low entropy model may require us to make additional
assumptions about the data, which is unreasonable due to the lack of support in
the prior knowledge of the data. In addition, making such unreasonable
assumptions would not guarantee that the resulting model is able to capture the
actual characteristics of the data. Thus, the only reasonable choice would be
avoiding such unreasonable assumptions and only relying on the given prior
information about the data although it would increase the entropy of the model
and make the model more uncertain about the data. This is exactly what the
Maximum Entropy principle addresses. Generally speaking, MaxEnt principle
identifies the best probability distribution, which maximizes the entropy, over
the dataset at hand given the prior knowledge about the data. The result MaxEnt
probability distribution uses the prior information optimally and best
summarizes the dataset, and is unbiased otherwise.

To be more specific, suppose we have a dataset $\boldsymbol{D}$ and a set of
functions $\mathcal{F} = \{f_i \mid f_i(\boldsymbol{D}) = S_i\}$ that compute
several statistics about the dataset $\boldsymbol{D}$. Such statistics will
serve as the prior information or background knowledge about the dataset. Using
this prior information as constraints, it defines a set of probability
distributions $\mathcal{P}$ that are consistent with the given dataset
statistics, e.g.\ $\mathcal{P} = \{p \mid \mathbb{E}_{p}[f_i(\boldsymbol{D})] =
S_i, \forall f_i \in \mathcal{F} \}$ where $\mathbb{E}_{p}[\cdot]$ denotes the
expectation under the probability distribution $p$. Among all these possible
probability distributions, MaxEnt principle identifies the distribution $p^{*}$
which maximizes the entropy,
\begin{align*}
	p^{*} = \argmax_{p \in \mathcal{P}} H(p)~.
\end{align*}
Here, $H(p)$ represents the entropy of probability distribution $p$.

\subsection{Subjective Interestingness}
\label{sec:pre_surprise}

In the data mining and knowledge discovery, the aim is to uncover the highly
informative information with respect to the prior knowledge or what we have
already known about the data --- we are not quite interested in what we already
know or what we can trivially infer from such knowledge.

To this end, we introduce the concept of subjective interestingness or
surprisingness. Suppose we have a probability distribution $p$ which models our
current beliefs about the data. When we evaluate the data mining results, we can
use $p$ to determine the likelihood of a result under our current knowledge
about the data. If the likelihood is high, this indicates that we probably
already know about this result, or we can easily infer such result. Thus,
reporting it would provide little novel information about the data. On the
contrary, if the likelihood is very low, the result could be very interesting or
surprising, which means it conveys a lot of new information compared to what we
have already known. In Section~\ref{sec:search}, we will formal define a
quantitative criterion to measure the subjective interestingness. In the rest of
this paper, we will assume that the two words interesting and surprising refer
to the same concept in our context and use them interchangeably.


\section{MaxEnt Model on Graph}
\label{sec:model}
In this section, we will define the MaxEnt model over graphs. Although there
exist several other generative models for graphs, e.g.\ the classic stochastic
block models~\cite{stochastic:block} which are well studied and widely used to
recover community structures in graphs, we choose to adopt the MaxEnt framework
here due to its natural fit for the interactive and iterative knowledge
discovery scenario in this paper.  Before we formally state the MaxEnt model, we
first describe some basic statistics about graphs, which will serve as the prior
information about the graph. Then, we will introduce the MaxEnt model over
graphs by applying the Principle of Maximum Entropy, and finally, we will
describe how we can estimate the graph MaxEnt model by maximizing the
likelihood.

\subsection{Notations for Graph Prior Knowledge}
\label{sec:v_degree}
In our scenario, we choose to use in-degrees and out-degrees of vertices and
subgraphs to characterize the prior knowledge of a given graph $G = (V, E)$. The
in-degrees and out-degrees of vertices are types of graph statistics that
describe the given graph from a global perspective. For the purpose of
convenience, we normalized the in-degree and out-degree of a vertex into the
range of $[0,1]$ by dividing the total number of vertices in the graph. In the
rest of this paper, we will use the terms in-degree and out-degree to refer to
the normalized in-degree and out-degree for each vertex. On the other hand,
subgraphs identify local information about graphs, which could be useful as
prior knowledge about the local structures of graphs. Although there are many
statistics available for subgraphs, we choose the densities of subgraphs to
characterize the local structures of the entire graph, which is defined as:
\begin{align*}
	f(G^{\prime}) = \frac{|E^{\prime}|}{|V^{\prime}|^2}, 
	~\text{where}~G^{\prime} = (V^{\prime}, E^{\prime}),~\text{and}~G^{\prime}
	\subseteq G,
\end{align*}
where $|V^{\prime}|$ and $|E^{\prime}|$ represent the number of vertices and
edges in the subgraph $G^{\prime}$, respectively. Here, notice that we are
considering a more general scenario that the edge from a vertex to itself is
allowed.

Let $d_{\mathit{in}}(v)$ and $d_{\mathit{out}}(v)$ to represent the in-degree
and out-degree of a vertex $v$, and $f(G^{\prime})$ denote the density of the
subgraph $G^{\prime}$. Suppose $\mathcal{G}$ is the space that contains all the
possible graphs that have $|V|$ vertices. Let $p$ be the probability
distribution defined over the graph space $\mathcal{G}$, then the expectation of
in-degree and out-degree of a given vertex $v$ and the expectation of the
density of a given subgraph $G^{\prime}$ would be:
\begin{align*}
	\mathbb{E}_{p}\left[ d_{\mathit{in}}(v) \right] & = \sum_{G \in \mathcal{G}}
	p(G) d_{\mathit{in}}(v) \\ 
	\mathbb{E}_{p}\left[ d_{\mathit{out}}(v) \right] & = \sum_{G \in \mathcal{G}}
	p(G) d_{\mathit{out}}(v) \\
	\mathbb{E}_{p}\left[ f(G^{\prime}) \right] & = \sum_{G \in \mathcal{G}}
	p(G) f(G^{\prime}) \\
\end{align*}

\subsection{MaxEnt Model with Prior Information}
\label{sec:maxent}

In this section, we will derive a global statistical model for graphs based on
the given graph prior knowledge. The graph prior information is provided in the
form of vertex degrees and the densities of various subgraphs as we
discussed in Section~\ref{sec:v_degree}. For a given graph $G = (V, E)$, suppose
we are given a set of vertex degree constraints $\mathcal{D} =
\{d_{\mathit{in}}(v_i) = D_i^{\mathit{in}}, d_{\mathit{out}}(v_i) =
D_i^{\mathit{out}} \mid v_i \in V\}$ and a set of subgraph density
constraints $\mathcal{F} = \{f(G^{\prime}) = F_{G^{\prime}} \mid G^{\prime}
\subseteq G\}$. Notice that the subgraph constraints $\mathcal{F}$ may not
necessarily contain every possible subgraph of $G$, which is also infeasible in
practice. We would like to infer a probability distribution $p$ over the
space of all possible graphs $\mathcal{G}$ that is consistent with information
given by the constraints $\mathcal{D}$ and $\mathcal{F}$. In other words, we
want to determine how likely is a graph $G \in \mathcal{G}$ given these
vertex degree and subgraph density constraints $\mathcal{D}$ and $\mathcal{F}$.

In order to derive a good statistical model, we adopt a principled and
statistically well-founded approach. We employ the MaxEnt principle introduction
in Section~\ref{sec:pre_maxent}. To formally define the MaxEnt distribution, we
first need to specify the space $\mathcal{P}$ that contains all the graph
probability distribution candidates which conform the given prior information.
Given the constraints $\mathcal{D}$ and $\mathcal{F}$ as prior knowledge, the
graph probability distribution space can be defined as: 
\begin{align*}
	\mathcal{P} = \{p \mid & \mathbb{E}_p[d_{\mathit{in}}(v_i)] =
	D_i^{\mathit{in}}, \mathbb{E}_p[d_{\mathit{out}}(v_i)] = D_i^{\mathit{out}},
	\\
	& \mathbb{E}_p[f(G^{\prime})] = F_{G^{\prime}}, \forall d_{\mathit{in}}(v_i),
	d_{\mathit{out}}(v_i) \in \mathcal{D}, \forall f(G^{\prime}) \in
	\mathcal{F}\}~.
\end{align*}
Among all these candidate distributions, we choose the distribution $p^{*}$
which maximizes the entropy $H(p)$. To infer the MaxEnt distribution, we rely
on a classical theorem in~\cite{csiszar:75:i-divergence} which states that a
distribution $p^{*}$ is the MaxEnt distribution if and only if it can be written
as an exponential form. In our scenario, the MaxEnt distribution would be:
\begin{align}
	p^{*}(G) \propto \exp \Bigg( & \sum_{d_{\mathit{in}}(v_i) \in \mathcal{D}}
	\lambda_i^{\mathit{in}} d_{\mathit{in}}(v_i) + \sum_{d_{\mathit{out}}(v_i)
	\in \mathcal{D}} \lambda_i^{\mathit{out}} d_{\mathit{out}}(v_i) \nonumber \\
	& + \sum_{f(G^{\prime}) \in \mathcal{F}} \lambda_{G^{\prime}}
	f(G^{\prime})\Bigg)~.
	\label{eq:3.2.1}
\end{align}
Here, $\lambda_i^{\mathit{in}}$, $\lambda_i^{\mathit{out}}$ and
$\lambda_{G^{\prime}}$ are the model parameters. 

By rearranging the terms within the summations, we could further factorize the
MaxEnt distribution $p^{*}$ into the product of a number of Bernoulli
distributions:
\begin{align*}
	p^{*}(G) = \prod_{v_i, v_j \in V} p^{*}(\mathbb{I}[(v_i, v_j) \in E]),
\end{align*}
where
\begin{align*}
	p^{*}(\mathbb{I}[(v_i, v_j) \in E] = 1) =
	\frac{\exp\left(\sum_{\lambda_i \in \Lambda} \lambda_i
	\right)}{\exp\left(\sum_{\lambda_i \in \Lambda} \lambda_i \right) + 1},
	\text{ or } 0,~1.
\end{align*}
Here, $\mathbb{I}[(v_i, v_j) \in E]$ is an indicator function which equals to
$1$ if the edge $(v_i, v_j)$ belongs to the edge set $E$ of the graph $G = (V,
E)$, otherwise $0$. $\Lambda = \{\lambda_{G^{\prime}} | \forall G^{\prime} =
(V^{\prime}, E^{\prime}), f(G^{\prime}) \in \mathcal{F}, v_i, v_j \in
V^{\prime}\} \cup \{\lambda_{i}^{\mathit{out}}, \lambda_{j}^{\mathit{in}}\}$ is
a set of model parameters $\lambda_{G^{\prime}}$ where the corresponding
subgraphs $G^{\prime}$ that are used to provide the prior information of the
graph $G$ contain the specific vertices $v_i$ and $v_j$, plus the model
parameters $\lambda_i^{\mathit{out}}$ and $\lambda_j^{\mathit{in}}$ for the
out-degree and in-degree constraints of the vertices $v_i$ and $v_j$,
respectively.

\subsection{MaxEnt Distribution Estimation}
\label{sec:maxent_est}

 \begin{algorithm}[t]
  \SetAlgoLined
  \SetKwInOut{Input}{input}
  \SetKwInOut{Output}{output}

  \Input{Graph $G = (V, E)$, vertex degree constraints $\mathcal{D} =
  \{d_{\mathit{in}}(v_i) = D_i^{\mathit{in}}, d_{\mathit{out}}(v_i) =
  D_i^{\mathit{out}} \mid v_i \in V\}$, and subgraph constraints
  $\mathcal{F} = \{f(G^{\prime}) = F_{G^{\prime}} \mid G^{\prime} \subseteq
  G\}$.}
  \Output{MaxEnt distribution $p^{*} \leftarrow p$.}
  \BlankLine
  $p \leftarrow$ a uniform distribution where $p(\mathbb{I}[(v_i, v_j) \in E] =
  1) = \frac{1}{2},~\forall~v_i,v_j \in V$\;
  \While{not converged}{
	  \For{$d_{\cdot}(v_i) \in \mathcal{D}$}{
		  $h \leftarrow \mathbb{E}_p[d_{\cdot}(v_i)]$, $\tilde{h} \leftarrow
		  D_i^{\cdot}$\;
		  $x \leftarrow \frac{\tilde{h}(1 - h)}{h (1 - \tilde{h})}$\;
		  $p(\mathbb{I}[(v_i, v_j) \in E] = 1) \leftarrow \frac{x \cdot
			  p(\mathbb{I}[(v_i, v_j) \in E] = 1)}{1 - (1 - x) \cdot
				  p(\mathbb{I}[(v_i, v_j) \in E] = 1)}$, for all $v_j \in V$\;
	  }
	  \For{$f(G^{\prime}) \in \mathcal{F}$}{
		  $h \leftarrow \mathbb{E}_p[f(G^{\prime})]$, $\tilde{h} \leftarrow F_{G^{\prime}}$\;
		  $x \leftarrow \frac{\tilde{h}(1 - h)}{h (1 - \tilde{h})}$\;
		  $p(\mathbb{I}[(v_i, v_j) \in E] = 1) \leftarrow \frac{x \cdot
			  p(\mathbb{I}[(v_i, v_j) \in E] = 1)}{1 - (1 - x) \cdot
				  p(\mathbb{I}[(v_i, v_j) \in E] = 1)}$, for all $v_i, v_j \in
				  V^{\prime}, G^{\prime} = (V^{\prime}, E^{\prime})$\;
	}
  }
  \Return{p}\;
  \caption{Iterative Scaling algorithm for estimating MaxEnt distribution over
  	graphs}
  \label{alg:3.3.1}
 \end{algorithm}

 In order to estimate the parameters of the MaxEnt distribution mentioned in
 the previous section, e.g.\ $\lambda_{G^{\prime}}$, $\lambda_i^{\mathit{in}}$
 and $\lambda_i^{\mathit{out}}$, we follow a standard approach and adopt the
 well-known Iterative Scaling (IS) algorithm~\cite{1972:iterative:scaling} to
 infer the MaxEnt model over graphs. Algorithm~\ref{alg:3.3.1} describes the
 detail of this IS algorithm where $d_{\cdot}(v_i)$ denotes either
 $d_{\mathit{in}}(v_i)$ or $d_{\mathit{out}}(v_i)$. Similarly, $D_i^{\cdot}$
 denotes either $D_i^{\mathit{in}}$ or $D_i^{\mathit{out}}$. Generally speaking,
 for each vertex degree constraint $d_{\cdot}(v_i) \in \mathcal{D}$ or subgraph
 density constraint $f(G^{\prime}) \in \mathcal{F}$, the IS algorithm updates
 the probability distribution $p$ such that the expectation of the vertex
 degree or subgraph density under distribution $p$ will be consistent with the
 given value in the corresponding constraint. Obviously, during such a single
 update, we may change the expected vertex degree or subgraph density
 corresponding to other constraints. Thus, several iterations are needed
 until the probability distribution $p$ converges. The proof of the convergence
 for the IS algorithm is beyond the scope of this paper. Readers who are
 interested in this topic, please refer to the Theorem 3.2
 in~\cite{csiszar:75:i-divergence}. In practice, it typically takes on the order
 of seconds for the IS algorithm to converge.

\section{Interactive Discovery of Connected Subgraphs}
\label{sec:search}

\begin{figure*}[!t]
	\centering
	\includegraphics[width=0.9\textwidth]{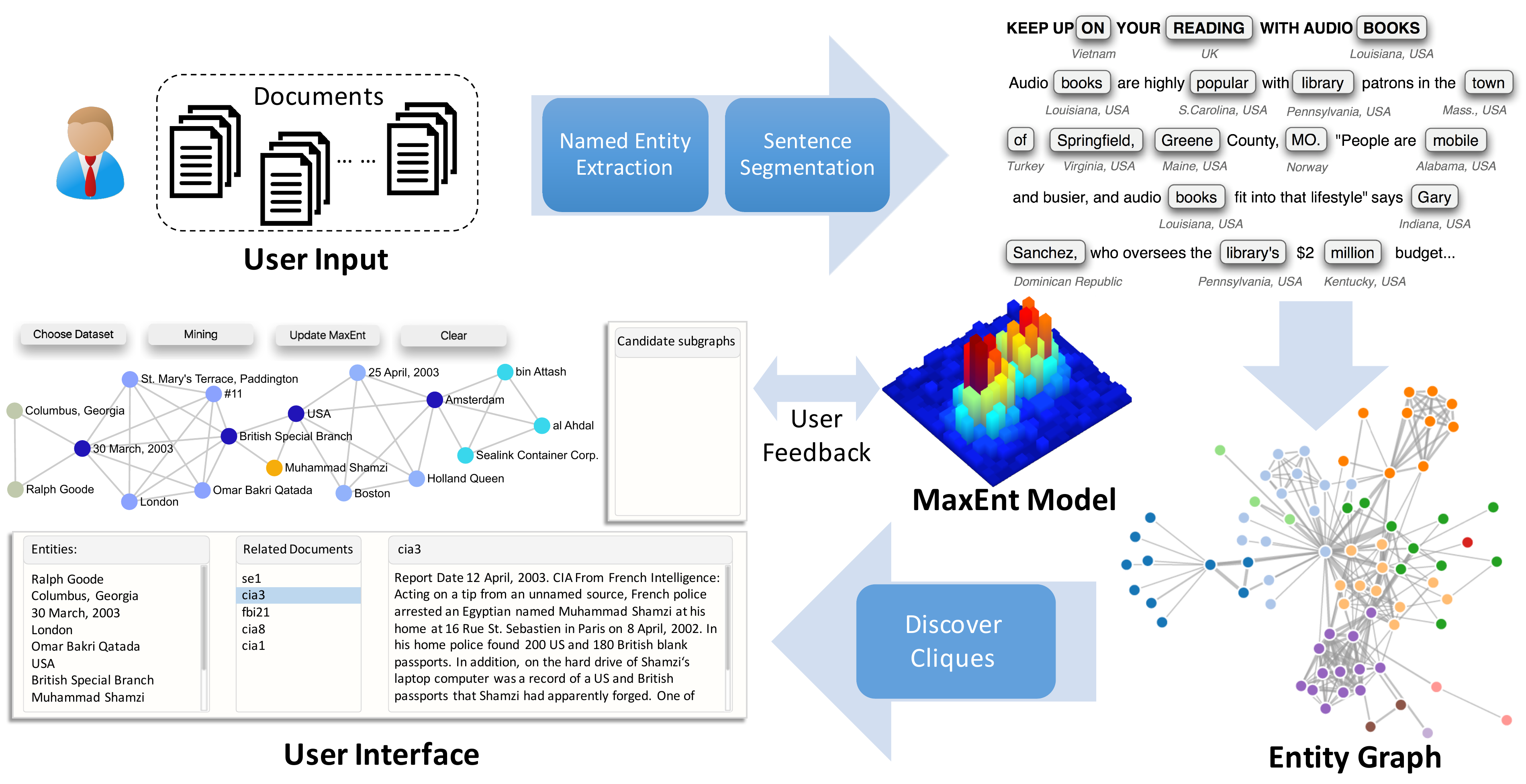}\vspace{-0.1in}
	\caption{The architecture of the proposed connected subgraph discovery
	framework. The user input is a text corpus with a collection of documents.
	After named entity extraction and sentence segmentation, an entity graph is
	created based on the entity co-occurrence on the sentence level. Then the
	graph MaxEnt model is inferred over the entity graph, and the subgraphs
	(cliques in our scenario) are identified from the entity graph. By
	iteratively interacting with the MaxEnt model through the user interface,
	interesting connected subgraph patterns could be discovered from the entity
	graph.}
	\label{fig:4.1.1}\vspace{-0.2in}
\end{figure*}

In this section, we illustrate our visualization framework that discovers
connected subgraphs from an entity graph with an interactive manner.
Figure~\ref{fig:4.1.1} shows the architecture of the proposed interactive
visualization framework. Although our proposed framework can be used to analyze
graphs originated from various types of raw data, e.g.\ biology data,
social networks, we focus on the entity graph constructed from text corpus in
this paper. In this scenario, the user input is a collection of text documents
from which the entity graph is constructed (Section~\ref{sec:graph}). With the
entity graph, we infer a background MaxEnt model (Section~\ref{sec:back}) and
define an interestingness measure (Section~\ref{sec:interest}), which will be
used to automatically discover connected subgraphs (Section~\ref{sec:greedy}) or
guide the user exploration process (Section~\ref{sec:user}).

\subsection{Creating Entity Graph}
\label{sec:graph}
Briefly speaking, we construct the entity graph based on the entity
co-occurrences on the sentence level in the text corpus. To be more specific,
given a text corpus $\mathit{TC} = \{\mathit{doc}_1, \mathit{doc}_2, \ldots,
\mathit{doc}_n\}$, we perform the sentence segmentation on each document
$\mathit{doc}_i \in \mathit{TC}$ to split each document into a set of sentences,
e.g.\ $\mathit{doc}_i = \{\mathit{sent}_{i,1}, \mathit{sent}_{i, 2}, \ldots,
\mathit{sent}_{i,m}\}$, and also extract named entities from each document
$\mathit{doc}_i$. 

With the extracted named entities and segmented sentences from the
text corpus, we can build the undirected entity graph $G = (V, E)$ with the
following approach. The vertex set $V$ of the graph is just the set of all the
extracted named entities. For the edge set $E$, if two named entities
$\epsilon_i$ and $\epsilon_j$ appear together in some sentence
$\mathit{sent}_{k,l}$ in the document $\mathit{doc}_k$, we add an undirected
edge $(\epsilon_i, \epsilon_j)$ into $E$. Although the entity graph created here
is undirected, as we mentioned in Section~\ref{sec:pre_graph}, an undirected
graph can be treated as a special case of a directed graph. Thus, our graph
MaxEnt formalization can be easily extended to the scenario of undirected
graphs.

\subsection{Background MaxEnt Model}
\label{sec:back}
Next, given the entity graph, we discuss how to specify the background MaxEnt
model which incorporates the basic prior knowledge about the given graph. In
order to discover non-trial connected subgraphs, we need some basic background
information about the entity graph and infer a background MaxEnt model so that
we can evaluate the surprisingness of patterns (connected subgraphs here) as we
discussed in Section~\ref{sec:pre_surprise}. In our scenario, we choose to use
the degree of each vertex in the graph as the prior knowledge to infer the
background MaxEnt model. Formally, for the entity graph $G = (V, E)$, the prior
information that used as constraints to infer the background MaxEnt model is
$\mathcal{D} = \{d(v_i) = D_i \mid v_i \in V\}$. Recall the form of the MaxEnt
distribution described in Equation~\eqref{eq:3.2.1}, the background MaxEnt model
$p_{\mathit{back}}$ in this scenario would be:
\begin{align*}
	p_{\mathit{back}}(G) \propto \exp\left(\sum_{d(v_i) \in \mathcal{D}}
	\lambda_i d(v_i) \right),
\end{align*}
and it can be inferred with the Iterative Scaling algorithm described in
Algorithm~\ref{alg:3.3.1}.

\subsection{Interestingness Measure}
\label{sec:interest}
In order to determine the interestingness or surprisingness of any subgraph
pattern $G^{\prime}$ discovered from the graph $G = (V, E)$, we propose an
interestingness measure that characterizes how much new information the subgraph
$G^{\prime}$ conveys with respect the background MaxEnt model
$p_{\mathit{back}}(G)$. We will do this by inferring two MaxEnt models, and
then compute the divergence between these two models. 

With the background MaxEnt model as described in Section~\ref{sec:back}, we need
another MaxEnt model which also incorporates the new information given by the
subgraph $G^{\prime}$ under consideration, e.g.\ the MaxEnt model inferred with
the prior information $\mathcal{D} = \{d(v_i) = D_i \mid v_i \in V\}$ and the
given subgraph density information $\{f(G^{\prime}) = F_{G^{\prime}}\}$. We call
such MaxEnt model $p_{G^{\prime}}$. Then, the interestingness measure we would
like to propose is defined as:
\begin{align}
	s(G^{\prime}) = \mathit{KL}(p_{G^{\prime}} || p_{\mathit{back}}).
	\label{eq:4.3.1}
\end{align}
Here, $\mathit{KL}$ denotes the Kullback-Leibler (KL)
divergence~\cite{cover2006elements} which is well studied, easy to compute, and
fits our modeling requirements. In theory, other divergence measures could also
be considered. With the KL divergence, larger $s(G^{\prime})$ indicates that
more new information is brought into the MaxEnt model $p_{G^{\prime}}$ by the
subgraph $G^{\prime}$, thus $G^{\prime}$ would be more interesting compared to
other subgraph patterns.

\begin{figure*}[!t]
	\centering
	\includegraphics[width=0.9\textwidth]{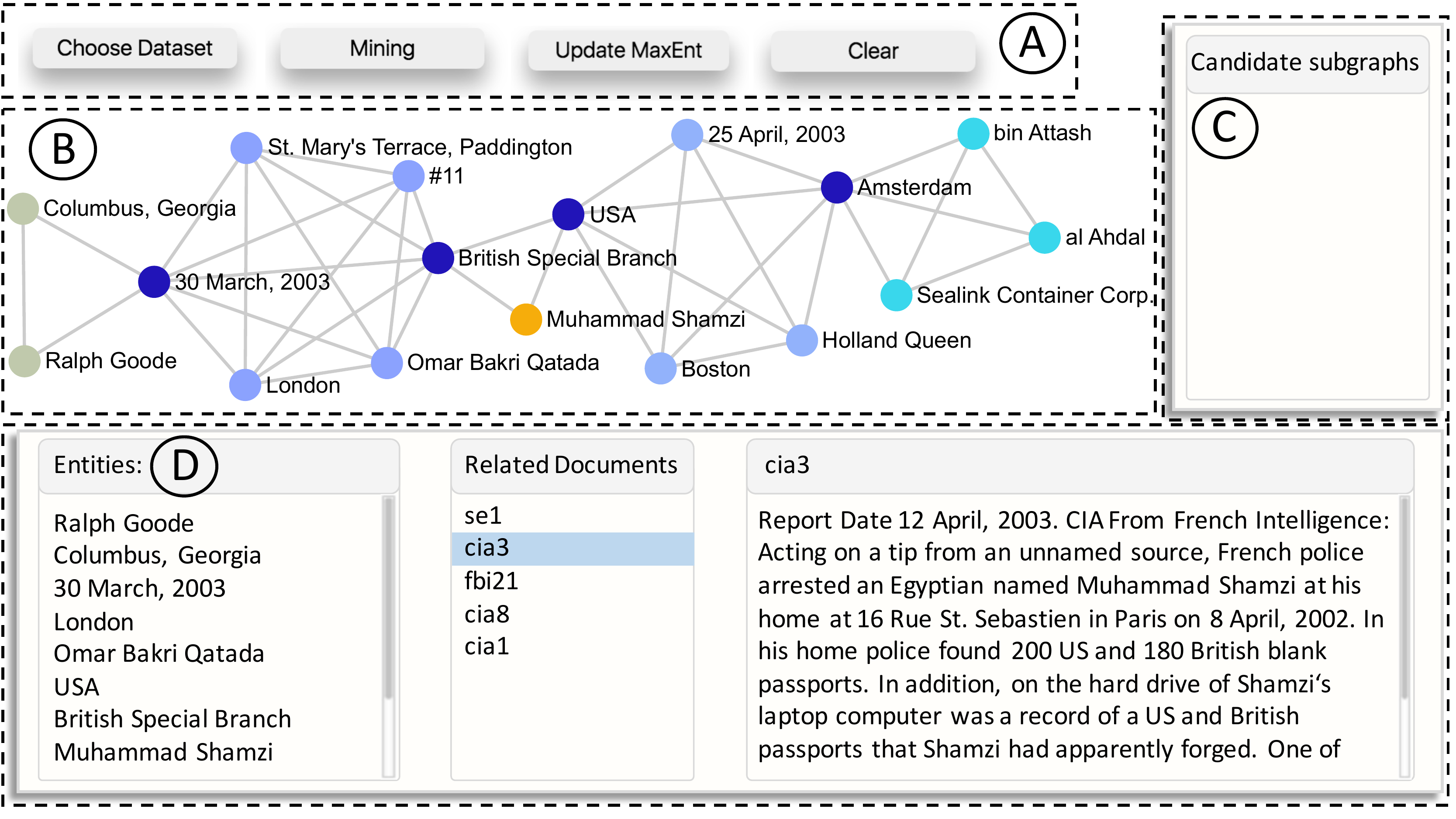}\vspace{-0.1in}
	\caption{The layout of the visualization interface. Region (A) contains the
	functional buttons where the user can load datasets, execute the automatic
	connected subgraph discovery, update the MaxEnt model, and clear the
	displayed results. Region (B) displays the current connected subgraph
	pattern under investigation. Region (C) lists the candidate subgraphs that
	can extend the displayed connected subgraph pattern in the user centric
	exploration process. Region (D) shows the corresponding entities and
	documents involved in the displayed connected subgraph pattern.}
	\label{fig:4.5.1}\vspace{-0.2in}
\end{figure*}

\subsection{Automatic Connected Subgraph Discovery}
\label{sec:greedy}
In this section, we describe a greedy heuristic strategy to automatically
discover interesting connected subgraph patterns from the entity graph $G$ with
the MaxEnt model and the interestingness measure defined above.  Ideally, to
discover a set of interesting connected subgraphs, we could exhaustively explore
the entire search space, find all the possible connected subgraphs, evaluate
their interestingness with the criterion defined in Section~\ref{sec:interest},
and choose the top $K$ candidates. However, the entire search space for the
connected subgraphs could be quite large when the subgraph patterns discovered
from the graph $G$ is huge. In this case, the intuitive exhaustive exploration
approach would be very inefficient or even infeasible in practice. Moreover, the
search space does not exhibit a particular structure which we could leverage to
perform an efficient search. Hence, we turn to heuristics.

\begin{algorithm}[!t]
 \SetAlgoLined
 \SetKwInOut{Input}{input}
 \SetKwInOut{Output}{output}
 \SetKwFunction{KwFnStart}{discoverSubgraphs}
 \SetKwFunction{KwFnEligible}{candidateSubgraph}
 \SetKwFunction{KwFnChain}{extend}
 \SetKwFunction{KwFnUpdateModel}{UpdateMaxEntModel}

 \Input{background MaxEnt model $p_\mathit{back}$; \\ entity graph $G = (V, E)$;
 		\\$K$, the desired number of connected subgraphs.}
 \Output{the set of connected subgraph $\mathcal{CS}$.}
 \BlankLine
 $\mathcal{CS} \leftarrow \emptyset$\;
 $\mathcal{G}^{\prime} \leftarrow$ \KwFnStart{$G$}\;\label{alg2:subgraph}
 \While{$(|\mathcal{CS}| < K)$}{
	 $\mathit{CS} \leftarrow \arg \max\limits_{G^{\prime} \in \mathcal{G}^{\prime}}
	 s(G^{\prime})$\;\label{alg2:start}
	 $\mathcal{G}_{c}^{\prime} \leftarrow $ \KwFnEligible{$\mathcal{G}^{\prime}$,
	 $\mathit{CS}$}\;\label{alg2:biCandidates}
	 \While{$|\mathcal{G}_c^{\prime}| \neq 0$}{\label{alg2:extend}
		 $G^{\prime}_n \leftarrow \arg \max \limits_{G^{\prime} \in
		 \mathcal{G}^{\prime}_c} s(G^{\prime})$\;
		 $\mathit{CS} \leftarrow $ \KwFnChain{$\mathit{CS}$, $G^{\prime}_n$}\;
		 $\mathcal{G}^{\prime}_c \leftarrow $
		 \KwFnEligible{$\mathcal{G}^{\prime}$, $\mathit{CS}$}\;
	 }\label{alg2:extend_end}
    $p_{\mathit{back}} \leftarrow $ \KwFnUpdateModel{$p_{\mathit{back}}$,
	$\mathit{CS}$}\; \label{alg2:updateModel}
    $\mathcal{CS} \leftarrow \mathcal{CS} \cup \{\mathit{CS}\}$\;
 }
 \Return{$\mathcal{CS}$}\;
 \caption{Greedy Heuristic for Connected Subgraph Discovery} \label{alg:4.4.1}
\end{algorithm}

We adopt a simple iterative greedy search strategy to automatically discover
interesting connected subgraph patterns. Algorithm~\ref{alg:4.4.1} illustrates
our proposed heuristic greedy search approach. Given an entity graph $G = (V,
E)$, we first discover a set of subgraph patterns from $G$
(Line~\ref{alg2:subgraph}). Various subgraph patterns have been studied in the
realm of graph mining, e.g.\ communities, dense subgraphs, cliques. In our text
corpus analysis scenario, we choose to use maximal cliques~\cite{Eppstein2010}
as our subgraph patterns. To discover a connected subgraph pattern, we start
from the most interesting clique with respect to the interestingness measure
defined in Equation~\eqref{eq:4.3.1} (Line~\ref{alg2:start}). We believe that
more interesting cliques are more likely to form an interesting connected
subgraph pattern, and thus help to reveal useful information in the entity graph
and in the corresponding text corpus. Then we extend the current connected
subgraph pattern $\mathit{CS}$ by greedily choosing the most interesting clique
from all the cliques that have at least one common vertex with the first or last
clique in the current pattern $\mathit{CS}$. This process continues until we
cannot find any other cliques to add into the current connect subgraph pattern
$\mathit{CS}$ (Line~\ref{alg2:extend}---\ref{alg2:extend_end}).  After we find a
connected subgraph pattern, we update the background MaxEnt model with the new
information in the discovered connected subgraph, e.g.\ each clique $G^{\prime}$
in this connected subgraph together with its density $f(G^{\prime})$, in order
to avoid discovering redundant patterns in the future iterations
(Line~\ref{alg2:updateModel}). The algorithm keeps discovering such surprising
connected subgraphs until the desired number of patterns are found or no more
connected subgraphs can be formulated.

\begin{figure*}[!t]
	\centering
	\includegraphics[width=0.88\textwidth]{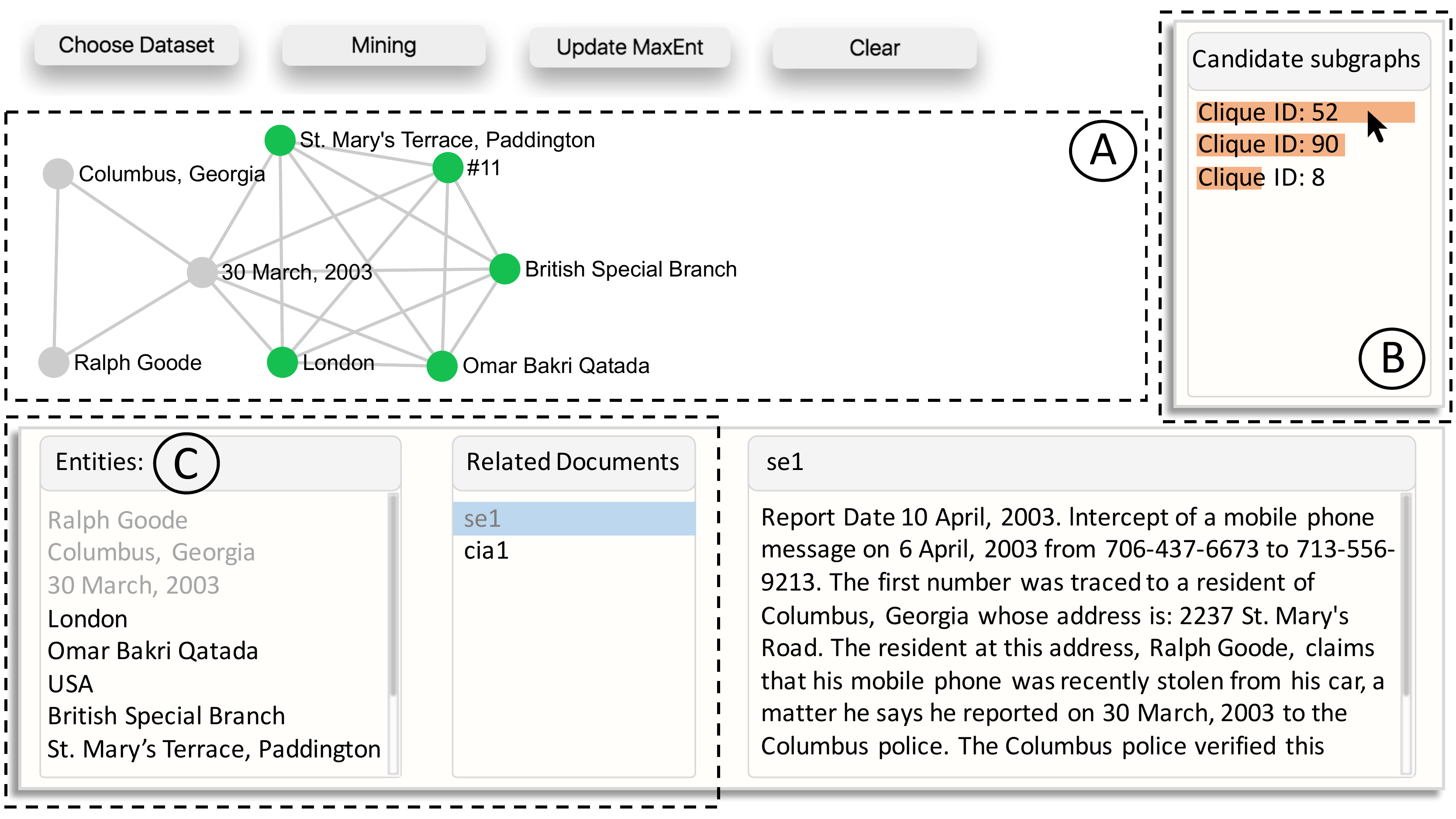}\vspace{-0.1in}
	\caption{A snapshot of the user centric exploration process. In the first
	step, the user chose the clique displayed in green color in Region (A). In
	the current step, three candidate subgraphs are listed in Region (B) with
	the most interesting one on the top based on the evaluation with the graph
	MaxEnt model. When the user selects one candidate subgraph from the list, it
	is displayed in Region (A) with gray color, and shows how the chosen
	subgraph is connected to the current pattern. At the same time, related
	entities and documents are also listed in Region (C) in gray color.}
	\label{fig:4.5.2}\vspace{-0.2in}
\end{figure*}

\subsection{User Centric Exploration}
\label{sec:user}
Although we have proposed automatic approach to discover the interesting
connected subgraph patterns from the entity graph as described in the last
section, we should not overlook the important role that user feedbacks play in
exploratory data mining tasks. In such scenarios, even if we have criteria to
evaluate the data mining results, we may still need a domain expert to verify
the results. If we enable an interactive scenario during the data mining
process, the data mining model can help the domain expert filter and refine the
large amount of patterns. While, on the other hand, the domain expert could also
verify the data mining results, update the data mining model, and lead the model
to the correct parts of the data to explore. That is the real power of
interactive, human-in-the-loop data mining approach.

Having realized the advantages of such human-in-the-loop data mining approach,
in this section, we proposed an interactive, user centric approach with a
visualization interface to discover the connect subgraph patterns from the
entity graph with the assistance of the MaxEnt model. Figure~\ref{fig:4.5.1}
shows the visualization interface that the analyst uses to communicate with the
MaxEnt model. Region (A) contains functionalities through which the user can
choose the dataset to explore (\textit{Choose Dataset} button), automatically
discover the connected subgraph patterns using the greedy heuristic discussed in
Section~\ref{sec:greedy} (\textit{Mining} button), update the MaxEnt model with
the discovered connected subgraph (\textit{Update MaxEnt} button), and clear the
results currently displayed (\textit{Clear} button). Region (B) displays the
current discovered connected subgraph pattern with color encodings. The vertices
within the same clique are displayed with the same color, and all the vertices
that connect adjacent cliques are also displayed using the same color, e.g.\
the vertices \textit{30 March, 2003}, \textit{British Special Branch},
\textit{USA}, and \textit{Amsterdam} in the displayed connected subgraph of
Figure~\ref{fig:4.5.1}. Region (C) shows a list of candidate subgraph patterns
that can be used to extend the current connected subgraph pattern. The different
candidate subgraph patterns in the list are identified by their unique IDs. In
the example shown in Figure~\ref{fig:4.5.1}, the displayed connected subgraph is
fully extended, thus this candidate subgraph list is empty. Finally, region (D)
displays the named entities involved in the current connected subgraph pattern
as well as the corresponding text documents in the corpus.

The user centric data exploration process works in the following way. By
clicking the button \textit{Choose Dataset}, the user can select the dataset he
wants to explore. By clicking the \textit{Mining} button, an automatic
discovery of the connected subgraph patterns will be performed using the greedy
heuristic search strategy as discussed in Section~\ref{sec:greedy}. In the user
centric exploration, the user is able to choose which subgraph he would like
to use to extend the current connect subgraph pattern. Figure~\ref{fig:4.5.2}
shows a snapshot of this user centric exploration process. Region (A) displays
the current incomplete connected subgraph pattern, and Region (B) lists all the
candidate subgraphs that can be used to extend this incomplete connected
subgraph pattern. Notice that the candidate subgraphs listed here are sorted
based on their interestingness measures (Equation~\eqref{eq:4.3.1}) evaluated by
the embedded MaxEnt model. The length of the orange bar indicates the
interestingness value of the corresponding candidate
subgraph~\cite{journals:tvcg:YalcinEB16}. The most interesting candidate
subgraph is listed on the top.  When the user moves the mouse over a specific
candidate subgraph in the list, it will be displayed the in Region (A) in gray
color showing how the chosen subgraph extends the current incomplete connected
subgraph pattern. The entities and text documents corresponding to the chosen
subgraph are also displayed with gray color in the entity list and related
document list in Region (C). However, the subgraph in such status is not
actually added into the current incomplete connected subgraph pattern. To add
the chosen subgraph, the user needs to click the specific candidate subgraph
listed in Region (B). By performing such click operation, the chosen candidate
subgraph will finally be used to extend the current incomplete connected
subgraph pattern, and be displayed in solid color as shown in
Figure~\ref{fig:4.5.1}. The corresponding candidate subgraph list, entity list
and related document list will also be updated.

\begin{figure*}[!t]
	\centering
	\includegraphics[width=0.9\textwidth]{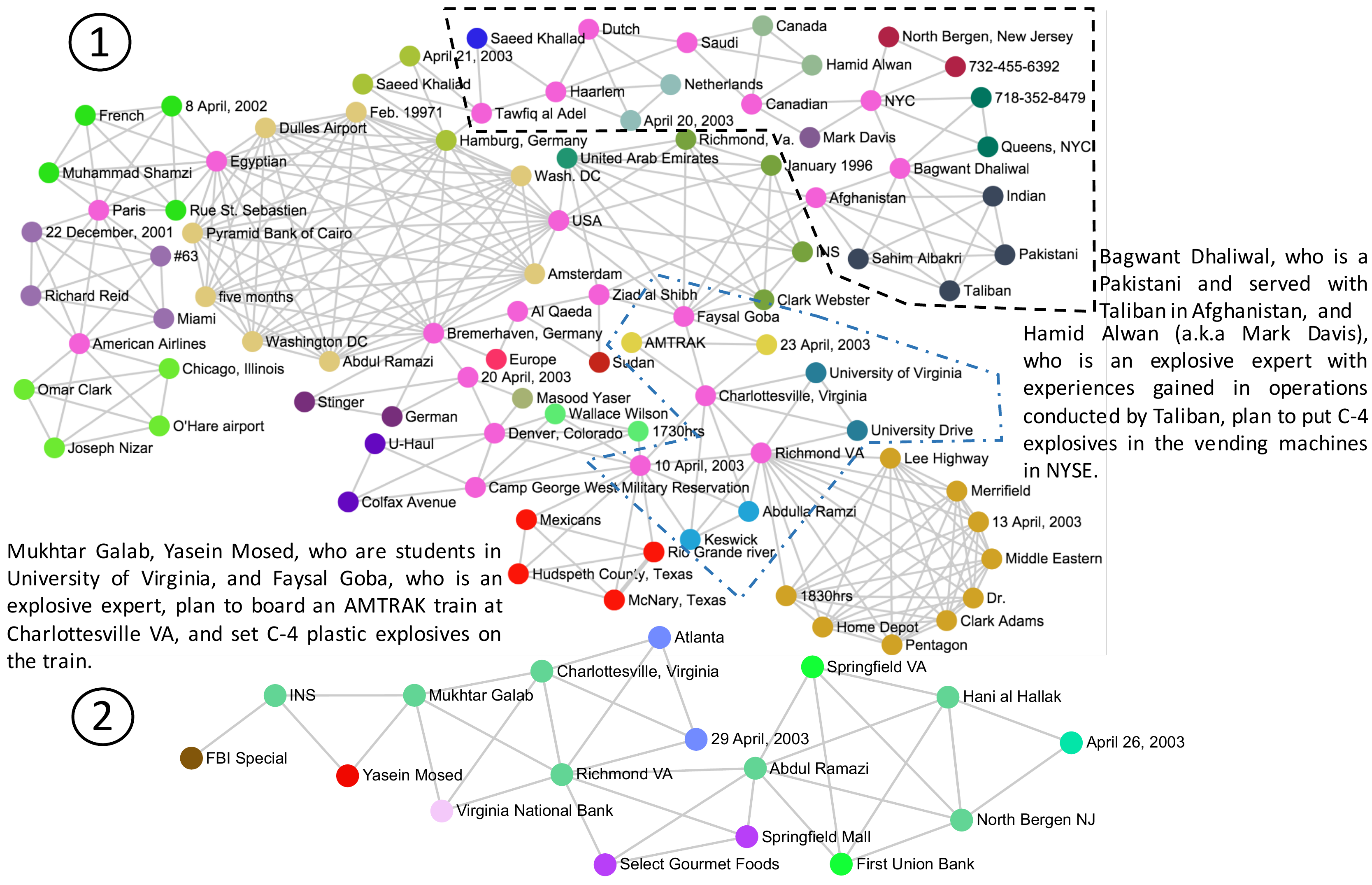}\vspace{-0.1in}
	\caption{Top connected subgraphs discovered by the greedy heuristic search
		strategy from \textit{Crescent} dataset. These two patterns help the
		intelligence analysts uncover two potential terrorist attacks from the
		dataset: (1) \textit{Bagwant Dhaliwal and Mark Davis plan to set
		explosives in the vending machines in NYSE}; (2) \textit{Mukhtar Galab
		et al.\ are going to blow up an AMTRAK train with C-4 plastic
		explosives}.}
		\label{fig:5.2.1}\vspace{-0.2in}
\end{figure*}

Whenever the user thinks the connected subgraph pattern he is exploring is
informative enough, or the current connected subgraph pattern is fully extended
(no other candidate subgraph can be added), the user can click the
\textit{Update MaxEnt} button to update the embedded graph MaxEnt model with the
discovered connected subgraph pattern. Such update operation is performed in the
same way, e.g.\ using each subgraph in the connected subgraph pattern and its
corresponding density, as we described in Section~\ref{sec:greedy} for automatic
connected subgraph discovery with the iterative greedy heuristic. By updating
the embedded graph MaxEnt model with the discovered connected subgraph pattern,
the new information is incorporated into the MaxEnt model so that the duplicate
patterns and redundant information will not be identified again in the future
iterations of the user centric exploration process over the same dataset.  By
clicking the \textit{Clear} button, the user can clear the results displayed for
the current iteration of the user centric exploration, and start a new iteration
of the analysis for the same dataset.  However, if the update MaxEnt operation
is not performed before this clear operation, the new information contained in
the last discovered connected subgraph pattern will not be incorporated into the
embedded graph MaxEnt model, and thus be disregarded.

\section{Experimental Results}
\label{sec:exp}

In this section, we describe the experimental results over some real world text
datasets, primarily text corpora from intelligence analysis domain. Although as
stated before, the proposed framework is widely applicable to many other types
of data where an entity graph can be created. We focus our experimental
investigations on answering the following questions:
\begin{enumerate}[itemsep=-1pt,topsep=-1pt]
	\item How to preprocess the text corpus so that the entity graph can be
		constructed? (Section~\ref{sec:dataset})
	\item How does the iterative greedy heuristic search strategy perform with
		respect to discovering interesting connected subgraphs when comparing to
		the true hidden plots of the text datasets? (Section~\ref{sec:result})
	\item How the visualization framework with embedded graph MaxEnt model could
		help and lead the user to identify the plots of the text corpus in the
		user centric exploration scenario? (Section~\ref{sec:case})
\end{enumerate}

\subsection{Dataset Processing}
\label{sec:dataset}
Given a text corpus, in order to construct the entity graph, we preprocess the
text documents in the corpus with natural language processing software from
Basis Technology~\cite{basis:nlp}. Sentence segmentation and named entity
extraction are performed on each document in the corpus to split each document
into a set of sentences and extract named entities from each document. A
co-reference operation is also performed over the extracted named entities so
that we can merge the named entities that refer to the same object into a single
unique entity. Then, we construct the entity graph based on the named entity
co-occurrences in sentences as we described previously in
Section~\ref{sec:graph} with igraph library~\cite{igraph:lib}.

\subsection{Results on Intelligence Datasets}
\label{sec:result}

\begin{figure*}[!t]
	\centering
	\includegraphics[width=0.9\textwidth]{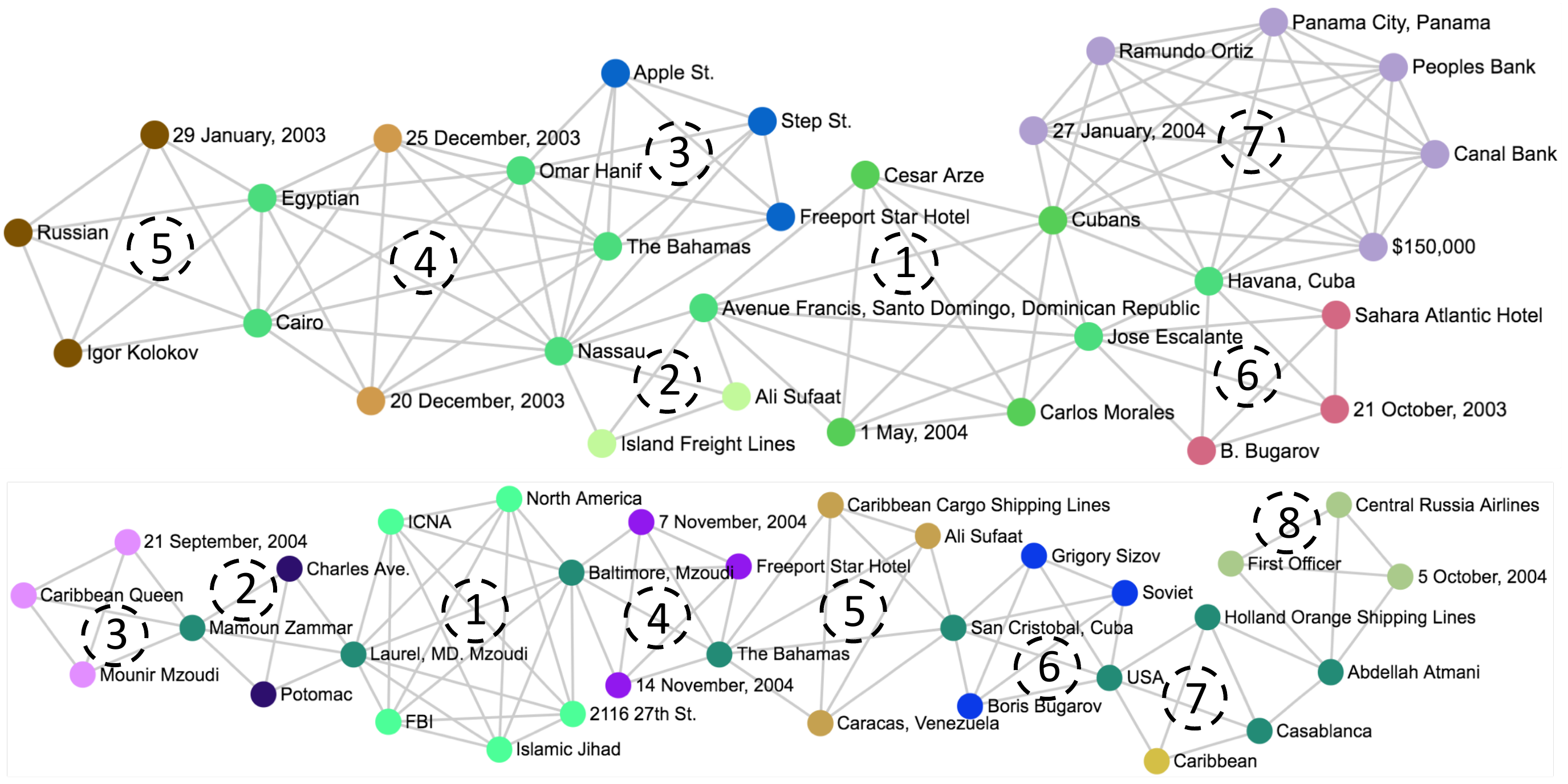}\vspace{-0.1in}
	\caption{Two connected subgraph patterns discovered by the user centric data
		exploration process from the \textit{Atlantic Storm} dataset. The
		numbers in dashed circles indicate the order that the connected subgraph
		pattern is extended following the interactions between the analyst and
		the graph MaxEnt model. These two patterns lead the analyst to reveal
		the hidden plot that \textit{Boris Bugarov et al.\ are trying to smuggle
		biological agents via Caribbean area to USA by using the Holland Orange
		Shipping Lines}.}
	\label{fig:5.3.1}\vspace{-0.2in}
\end{figure*}

In this section, we show the results of adopting the proposed iterative greedy
heuristic search strategy described in Section~\ref{sec:greedy} to a real world
intelligence analysis dataset \textit{Crescent}~\cite{Hughes:Discovery}. By
providing a collection of intelligence documents, the task for the
\textit{Crescent} dataset is to try to discover any imminent threats or possible
terrorist attacks by analyzing the intelligence documents. Through searching the
entity graph with the greedy heuristic strategy, a large connected subgraph
pattern which is the most interesting one with respect to the background MaxEnt
model defined in Section~\ref{sec:back} is discovered (pattern \ding{172} in
Figure~\ref{fig:5.2.1}). By reading through the corresponding intelligence
documents, we notice that the top right part of this connected subgraph pattern
(surrounded by the black dash line) helps us identify several pieces of useful
evidence: 
{\it\begin{itemize}[itemsep=-1pt,topsep=-1pt]
	\item Mark Davis, who works at Empire State Vending Service (ESVS),
		services the vending machines at New York Stock Exchange (NYSE). He
		is also known as Hamid Alwan, a Suadi national who received
		explosive training in Sudan and Afghanistan with Taliban.
	\item Bagwant Dhaliwal, who lives together with Mark Davis in Queens, NYC
		with a phone number 718-352-8479, is employed by ESVS. He was discovered
		that he fought with Taliban from 1990 to 1992.
	\item Hani al Hallak manages a carpet store in North Bergen, NJ with a phone
		number 732-455-6392. A fire happened at Hani al Hallak's carpet shop
		where C-4 explosives were discovered in the basement.
	\item Several calls were made from the number 718-352-8479 to 732-455-6392.
		In the most recent call, the caller said he would pick up the carpet on
		25 April, 2003.
\end{itemize}}
\noindent By connecting such evidence together, we are probably going to draw
the conclusion that an potential terrorist attack to the \textit{New York Stock
Exchange} (\textit{NYSE}) will happen soon.

This connected subgraph pattern help us uncover one of the plots hidden in this
\textit{Crescent} dataset. By updating the background MaxEnt model with this
pattern, we incorporate the information conveyed by this pattern into the
background MaxEnt model. After the update, the greedy heuristic search algorithm
identify another interesting connected subgraph pattern as shown by the pattern
\ding{173} in Figure~\ref{fig:5.2.1}. By checking corresponding intelligence
documents in the dataset, we discover the following important pieces of
evidence:
{\it
	\begin{itemize}[itemsep=-1pt,topsep=-1pt]
		\item Mukhtar Galab and Yasein Mosed who have been enrolled at
			University of Virginia was discovered holding expired student visas
			, and they have not attended any classes for the past two semesters.
		\item Mukhtar Galab, Yasein Mosed and Faysal Goba reserved three one-way
			first class tickets for AMTRAK Train \#19 from Charlottesville, VA
			to Atlanta, GA on 29 April, 2003. In addition, Faysal Goba received
			explosive training in Sudan in 1994 with Al Qaeda.
		\item Abdul Ramazi sent money of \$13,000 and \$8,500 to Mukhtar Galab
			and Hani al Hallak of North Bergen, NJ respectively. Muhammed bin
			Harazi served with Taliban from 1987 to 1993, and entered USA in
			March, 1993 with an alias name Abdul Ramazi.
	\end{itemize}}
\noindent Remember that one piece of evidence revealed by the previous connected
subgraph is \textit{Hani al Hallak manages a carpet store in North Bergen, NJ,
and a fire happend at his carpet shop where C-4 explosives were discovered in
the basement}. Combining with this evidence, we could connect these dots
together, and make the hypothesis that Abdul Ramazi purchases C-4 explosives
through Hani al Hallak and send the explosives to Mukhtar Galab, Yasein and
Faysal Goba who will probably set these explosives on the AMTRAK Train \#19 from
Charlottesville, VA to Atlanta, GA on 29 April, 2003.

Here, we should emphasize that by incorporating the new information contained in
the first discovered connected subgraph pattern (pattern \ding{172} in
Figure~\ref{fig:5.2.1}) into the background MaxEnt model, we successfully
identify another connected subgraph pattern which leads to a second potential
terrorist attack hidden in the dataset. This indicates that the newly discovered
pattern contains almost no redundant information and is surprising with respect
what we already know. Such results demonstrate that our proposed graph MaxEnt
model and greedy heuristic search strategy combined together are able to
discover interesting and surprising connected subgraph patterns from entity
graphs, which additionally leads the analysts to identify the plots that may
hide in the dataset. On the other hand, by iteratively identifying interesting
connected subgraph patterns and updating the graph MaxEnt model with the
discovered patterns, we can incorporate the knowledge we have just learned into
the MaxEnt model, and guarantee to discover no redundant information in the
future iterations of knowledge discovery. That's the most fascinating part of
iterative data mining with MaxEnt models in the scenario of data exploration and
knowledge discovery, identifying only interesting results with respect to what
we have already learned and at the same time reducing the redundancy as much as
possible.

\subsection{Use Case Study}
\label{sec:case}

In this section, we demonstrate the user centric data exploration with our
proposed graph MaxEnt model guided visualization framework by showing a use case
study on another real world intelligence dataset \textit{Atlantic Storm}. The
goal of the \textit{Atlantic Storm} dataset is to try to identify any potential
illegal international weapon transportation from a given collection of
intelligence documents. Figure~\ref{fig:5.3.1} shows two connected subgraph
patterns discovered by the analyst through the user centric data exploration
process. 

The analyst starts the exploration process by loading the \textit{Atlatic Storm}
dataset into the visualization framework. Then, from the entity graph
constructed from the \textit{Atlantic Storm} dataset, the system discovers all
the maximal cliques from the entity graph, and assigns an unique ID to each of
the maximal clique. By looking through a few maximal cliques ranked on the top
by the graph MaxEnt model and the corresponding intelligence documents, the
analyst chooses a maximal clique with six vertices (the clique marked with
number 1 in dashed circle in the top connected subgraph pattern in
Figure~\ref{fig:5.3.1}) to start his exploration of the \textit{Atlantic Storm}
dataset since it contains three person entities and one location entity.
Usually in intelligence analysis, these two types of entities play an important
role in uncovering the hidden plots. After choosing the starting maximal clique,
the system updates the candidate maximal cliques that can be used to extend
the current connected subgraph pattern, and display them in the user
visualization interface for the analyst to select. This time, the analyst
chooses the maximal clique with four vertices (the clique marked with number 2
in dashed circle) to extend the current connected subgraph pattern. By repeating
these steps, the analyst successfully extends the current pattern to contain
seven maximal cliques discovered from the entity graph. Although this connected
subgraph pattern can still be further extended, the analyst thinks the remaining
candidate maximal cliques are not informative enough to be added into the
current pattern. Thus, he decides to end the current iteration of exploration
and update the graph MaxEnt model with this connected subgraph pattern (the top
connected subgraph pattern shown in Figure~\ref{fig:5.3.1}). However, with this
single connected subgraph pattern, the analyst thinks the evidence provided by
this pattern is not enough to draw any conclusion, thus he decides to start a
new iteration of exploration.

By repeating the exploration process described in the last paragraph, the
analyst identifies another connected subgraph pattern as shown at the bottom of
Figure~\ref{fig:5.3.1}, which contains eight maximal cliques (the numbers
indicate the order by which the pattern is extended). Notice that the connected
subgraph pattern at the bottom contains several entities that also appear in
the pattern on the top, e.g.\ \textit{Ali Sufaat}, \textit{The Bahamas}, and
\textit{Boris Bugarov} (refer to the same person with \textit{B.  Bugarov}),
which indicates that these two connected subgraph patterns may describe the same
hidden plot of the dataset. By reviewing the intelligence documents related to
these two connected subgraph patterns shown in Figure~\ref{fig:5.3.1} and
connecting the pieces of evidence together, the analyst identifies the following
possible illegal biological agent transportation:
\begin{quote}
	{\it Boris Bugarov and Jose Escalante, with the help of Abdellah Atmani who
		works for Holland Orange Shipping Lines, coordinate with each other to
		recruit Al Qaeda field agents to transport biological agents to
		Caribbean area via Holland Orange Shipping Lines. Jose Escalante, Cesar
		Arze and Carlos Morales are involved in transferring biological agents
		from Bahamas to USA.
	}
\end{quote}

Through this use case study, such results demonstrate that with the embedded
graph MaxEnt model in the visualization framework, good candidate subgraphs are
provided to the analyst in the user centric data exploration process so that
interesting connected subgraph patterns can be identified, which further serves
as informative hints to lead to the hidden plots in the intelligence datasets.

\section{Related Work}
\label{sec:related}
\textbf{MaxEnt models} have drawn much attention recently in the pattern mining
community, especially on the topic of mining representative/succinct/surprising
patterns, e.g.~\cite{Kiernan:2009:CCS:1631162.1631169}, as well as explicit
summarization~\cite{conf:sdm:DavisST09,mampaey:12:mtv}.~\citet{wang:06:summaxent}
summarized a collection of frequent patterns by using a row-based MaxEnt model,
heuristically mining and adding the most important itemsets in a level-wise
fashion.~\citet{tatti:06:computational} showed that querying such a model is
PP-hard.~\citet{mampaey:11:tell} gave a convex, MaxEnt model based heuristic,
allowing more efficient search for the most informative set of
patterns.~\citet{debie:11:dami} systematically formalized how to model a binary
matrix by the MaxEnt principle using row and column margins as background
knowledge.~\citet{tatti:12:apples} compared the informativeness of data mining
results given by different approaches over the same data by applying the binary
MaxEnt model.~\citet{Spyropoulou:local:patterns, DBLP:SpyropoulouBB14,
Spyropoulou:2011:IMP:2117684.2118182} formally defined Maximal Complete
Connected Subset (MCCS) patterns, and proposed to use the K-partite graph and
the MaxEnt model to discover surprising MCCS patterns from multi-relational data
with $n$-ary relationships.~\citet{konto:11:icdm,konto:13:numaxentit} introduced
a real-valued MaxEnt model and proposed a subjective interestingness measure
called \textit{Information Ratio} to iteratively discover the interesting
structures in real-valued data.

\textbf{Iterative data mining} was first introduced
by~\citet{hanhijarvi:09:tell}. The basic idea is to iteratively identify the
results from the data, which are most significant given our accumulated
knowledge about the data.  To assess significance, they built upon the
swap-randomization approach of~\citet{gionis:07:assessing} and evaluated
empirical p-values.~\citet{konto:13:numaxentit} and~\citet{mampaey:12:mtv}
demonstrated separately that ranking data mining results with a static MaxEnt
model leads to redundancies among the high-ranked patterns, and the iterative
data mining methodology provides a principled approach to keep the data mining
model updated and thus avoid such type of redundancy.

\textbf{Storytelling} or \textbf{finding plots} from text corpus is another
exploratory data mining task that has been extensively studied in recent a few
years. By finding a chain of intermediate articles that are maximally coherent
given either a start or end-point article,~\citet{shahaf-guestrin-journal}
studied the problem of summarizing a large collection of news articles by
identifying a chain of main events. Storytelling algorithms
\cite{Hossain:2012:SEN:2339530.2339742,storytelling-tkde,connectingpubmed}
provide algorithmic frameworks to automatically connect pieces of evidence
which may scatter into various different text documents and reveal the stories
hidden in the dataset.~\citet{Wu:2014:UPD:2664051.2664089} proposed a MaxEnt
based framework to identify the plots by detecting non-obvious coalitions of
entities from multi-relational datasets and further support iterative,
human-in-the-loop, knowledge discovery.~\citet{Ning:2015:UNR:2808797.2809329}
adopted a document similarity based storytelling algorithm to discover story
chains from news articles, and studied the relationships between the identified
story chains and the tweeters that belong to the same topics.

\textbf{Group structure visualization in graphs} has been a well studied
research topic in the visualization community. A lot of techniques have been
proposed to visualize the group structures in the node-link diagram
representation of graphs. Colors are often used to explicitly visualize the
group membership of vertices. Vertices that belong to only a single group are
simply colored with a unique color to represent the
membership~\cite{Dunne:2013:MSI:2470654.2466444}. For the scenario that vertices
can belong to multiple groups, one approach is to represent vertices by pie
charts with sections filled by corresponding colors that denote the groups the
vertices belong to~\cite{4906846,7042494,6295796}.~\citet{6634179} proposed to
use different size of the pie chart sections to encode the fuzzy membership
degrees, while for crisp overlapping
communities,~\citet{Xu:2013:VAS:2600534.2600544} used glyphs to encode group
overlap by integrating various visual channels, e.g.\ intensity of color, hue,
size, and shape. Some other approaches optimize the color assignment to maximize
the color differences between the adjacent neighbor
groups~\cite{5567116,6295791}. The group structures could also be visualized
using contours within the graph, e.g.\ vertices within the same contour belong
to the same group. The contour shape could be rectangles~\cite{5613447},
circles~\cite{4015433}, convex hulls~\cite{6658746}, and arbitrary
two-dimensional curves or
splines~\cite{6658746,JGAA-315,CGF:CGF3080,Dinkla2014}. However, group structure
visualization in graphs is not our primary focus here in this paper. Readers who
are interested in this topic would find a comprehensive survey of related work
in~\cite{eurovisstar.20151110}.

\section{Conclusion}
\label{sec:conclusion}
In this paper, we formally introduce the graph MaxEnt model, which is a
significant step to bring the well-found Maximum Entropy principle into the
graph data. With the proposed graph MaxEnt model, we study the problem of
discovering surprising connected subgraph patterns from entity graphs, which
could help to uncover interesting facts hidden inside the graphs. By designing a
MaxEnt model embedded visualization framework, we illustrate how the model
guided, human-in-the-loop iterative data mining process can help the exploratory
data mining tasks. Although we primarily focus on demonstrating our proposed
approach by showing the results over text datasets from intelligence analysis,
the theories and methods we present here are also applicable to graphs
originated from other types of data in general, e.g.\ biology and social network
data. Possible directions for future work may include improving the efficiency
of MaxEnt model estimation and the design of visualization interface, e.g.\
adopting better color encoding and layout algorithm to display the connected
subgraph patterns, to better support the interactive visual analytics of large
graphs.


\bibliographystyle{abbrvnat}
\bibliography{reference}

\end{document}